\documentclass[preprint,12pt]{elsarticle}

\usepackage{amssymb}

\usepackage{lineno}
\usepackage{hyperref}
\usepackage{ulem}

\biboptions{sort&compress}

\usepackage{float}
\usepackage{fullpage}

\restylefloat{table}

\journal{arXiv}

\begin{document}

\begin{frontmatter}

\newpage



\title{Version 2.0.0 - M-SPARC: Matlab-Simulation Package for Ab-initio Real-space
Calculations}


\author[label1]{Boqin Zhang}
\author[label1]{Xin Jing}
\author[label1]{Shashikant Kumar}
\author[label1]{Phanish Suryanarayana\corref{cor1}}
\ead{phanish.suryanarayana@ce.gatech.edu}
\cortext[cor1]{corresponding author}
\address[label1]{College of Engineering, Georgia Institute of Technology, Atlanta, GA 30332, USA}

\begin{abstract}
M-SPARC is a \textsc{Matlab} code for performing ab initio Kohn–Sham Density Functional Theory simulations. Version 2.0.0 of the software further extends its capability to include relativistic effects, dispersion interactions, and advanced semilocal/nonlocal exchange-correlation functionals. These features significantly increase the fidelity of first principles calculations that can be performed using M-SPARC. 

\end{abstract}


\end{frontmatter}

\section*{Metadata}
\label{metadata}
\vspace{-4.5mm}
\begin{table}[H]
\begin{tabular}{|l|p{6.5cm}|p{6.5cm}|}
\hline

\hline
C1 & Current code version & v2.0.0 \\
\hline
C2 & Permanent link to code/repository used for this code version &  \url{https://github.com/SPARC-X/M-SPARC}  \\
\hline
C3 & Code Ocean compute capsule & N/A \\
\hline
C4 & Legal Code License   & GNU General Public License v3.0 \\
\hline
C5 & Code versioning system used & git\\
\hline
C6 & Software code languages, tools, and services used & MATLAB 2013+ \\
\hline
C7 & Compilation requirements, operating environments \& dependencies & OS: Unix, Linux, MacOS, or Windows\\
\hline
C8 & If available Link to developer documentation/manual & \url{https://github.com/SPARC-X/M-SPARC/tree/master/doc} \\
\hline
C9 & Support email for questions & phanish.s@gmail.com\\
\hline
\end{tabular}
\end{table}

\vspace{-4.5mm}

\section*{Refers to} \vspace{-2mm}
Xu, Q., Sharma, A., and Suryanarayana, P., 2020. M-SPARC: MATLAB-Simulation Package for Ab-initio Real-space Calculations. SoftwareX, 11, 100423. \url{https://doi.org/10.1016/j.softx.2020.100423}



\newpage

\section{Description of the software-update}
\label{Description of the software-update}

M-SPARC \cite{xu2020m} is an electronic structure code written in  \textsc{Matlab} that is based on real-space finite-difference method. It can perform spin-unpolarized and polarized ab initio calculations based on pseudopotential Kohn–Sham Density Functional Theory (DFT) \cite{kohn1965self, hohenberg1964inhomogeneous} for extended systems such as nanotubes/nanowires,  surfaces, and crystals, as well as isolated systems such as molecules. In particular, the code can perform electronic ground state calculations for fixed atomic positions and cell dimensions (i.e., single-point calculations), geometry optimizations with respect to either atomic positions or cell volume, and microcanonical ensemble (NVE) molecular dynamics simulations, while employing norm-conserving pseudopotentials \cite{hamann2013optimized, troullier1991efficient}. In so doing, it can calculate the free energy, Hellmann–Feynman atomic forces, and Hellmann–Feynman stress tensor. 

M-SPARC can be regarded as the \textsc{Matlab} implementation of the large-scale parallel C/C++ code SPARC \cite{xu2021sparc, ghosh2017sparc2, ghosh2017sparc1}, with both codes employing the same algorithms, structure, input, and output. M-SPARC not only provides a suitable avenue for the  first principles investigation of systems with small/moderate size, but also provides a prototyping platform that allows for the rapid development and testing of new methods/algorithms in real-space DFT \cite{suryanarayana2013coarse, pratapa2016spectral, banerjee2016cyclic, xu2018discrete, kumar2020preconditioning, sharma2018real, codony2021transversal, diaz2021implementation, sharma2022real}, given the significant complexity of large-scale parallel codes  written in lower level programming languages such as C/C++ and Fortran. Indeed, the development of KSSOLV \cite{jiao2022kssolv} (\textsc{Matlab}) and DFTK \cite{herbst2021dftk} (\textsc{Julia}) for planewave DFT \cite{Martin2020}, and RESCU \cite{michaud2016rescu} (\textsc{Matlab}) and RSDFT \cite{RSDFT} (\textsc{Matlab}) for real-space DFT have  been  similarly motivated.

Version 2.0.0 of the M-SPARC software further extends its capability to include relativistic effects, dispersion interactions, and advanced exchange-correlation functionals beyond the generalized gradient approximation (GGA) \cite{Martin2020}, as described below. \vspace{-1mm}

\begin{itemize}
\item \emph{Spin-orbit coupling (SOC)}: SOC is a relativistic effect that refers to the coupling between the electron's orbital angular momentum and its spin angular momentum \cite{Martin2020}. It becomes increasingly prominent for heavier atoms, i.e., those with larger atomic numbers, and is known to play a significant role in determining their electronic structure.  M-SPARC incorporates SOC through relativistic norm-conserving pseudopotentials \cite{kleinman1980relativistic}, as implemented within the real-space method \cite{naveh2007real}.  \vspace{-0.5mm}

\item \emph{Dispersion interactions}: Van der Waals (vdW) interaction is a correlation effect that arises due to the coupling in electronic charge fluctuations between different parts of the system \cite{Martin2020}. This long-range dispersion interaction becomes increasingly important as the system gets more sparse, i.e., the inter-particle separation becomes larger. M-SPARC incorporates these interactions through the DFT-D3 correction \cite{grimme2010consistent} and the nonlocal vdW-density functional (vdW-DF)  \cite{dion2004van}, as implemented using the method proposed in Ref.~\cite{roman2009efficient}.  \vspace{-0.5mm}

\item \emph{Meta-GGA  functionals}: Meta-GGA exchange-correlation functionals represent the third rung of Jacob's ladder, i.e., one rung above GGA \cite{Martin2020}. Indeed, the sophistication and accuracy of the functionals increases as one goes up the ladder. In particular, in addition to the electron density and its gradient  used to define GGA, the kinetic energy density  is included in meta-GGA functionals. M-SPARC incorporates  meta-GGA through the SCAN functional \cite{sun2015strongly}, which satisfies all seventeen constraints known on the exact exchange-correlation functional. \vspace{-0.5mm}

\item \emph{Hybrid exchange-correlation functionals}: Hybrid exchange-correlation functionals represent the fourth rung of Jacob's ladder, i.e., one rung above meta-GGA, and therefore two rungs above GGA \cite{Martin2020}. In particular, in addition to the semilocal GGA/meta-GGA terms, a fraction of the Hartree-Fock exact exchange energy --- quantity that depends explicitly on the occupied orbitals --- is included in the exchange-correlation functional. M-SPARC incorporates exact exchange through the PBE0 \cite{adamo1999toward} and HSE \cite{heyd2003hybrid} functionals, as implemented using the methods proposed in Refs.~\cite{lin2016adaptively, spencer2008efficient, gygi1986self}. \vspace{-0.5mm}

\end{itemize}
In addition to the above, nonlinear core correction (NLCC) --- accounts for the nonlinearity in the exchange-correlation potential within pseudopotential generation --- has been implemented, a comprehensive testing framework with a large variety and number of examples has been developed, and the table of SPMS \cite{spms}   pseuodpotentials --- transferable and soft optimized norm-conserving Vanderbilt (ONCV) pseudopotentials \cite{hamann2013optimized} with NLCC  for the PBE \cite{perdew1996generalized} variant of the GGA exchange-correlation functional   --- has been  incorporated into the M-SPARC distribution. 

\begin{figure} [H]
\centering
\includegraphics[width=0.99\textwidth]{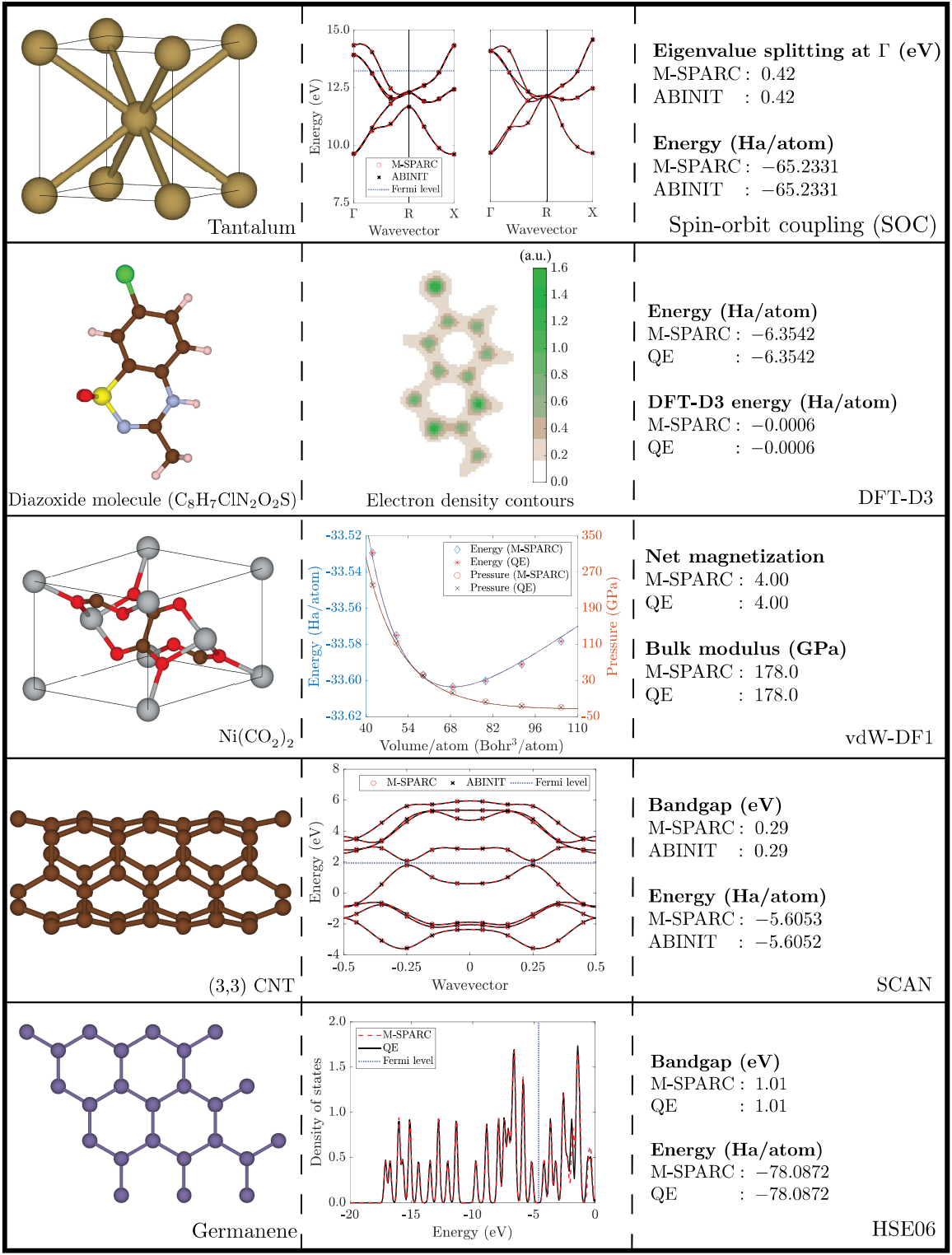}
\caption{Examples demonstrating the major new functionalities of M-SPARC v2.0.0.}
\label{Fig:Examples}
\end{figure}

We now demonstrate the aforementioned major new functionalities of M-SPARC through representative examples. Specifically, we consider (i) 2-atom primitive cell of  body-centered cubic (bcc) tantalum with PBE exchange-correlation, SOC through the relativistic ONCV pseudopotential from the PseudoDOJO set \cite{van2018pseudodojo}, 6$\times$6$\times$6 grid for Brillouin zone integration, and mesh size of 0.14 Bohr; (ii) diazoxide molecule with PBE exchange-correlation, dispersion interactions through DFT-D3, and mesh-size of 0.24 Bohr; (iii) 14-atom cell of bulk Ni(CO$_2$)$_2$ with  PBE exchange-correlation, dispersion interactions through vdW-DF, spin polarization, 2$\times$2$\times$2 grid for Brillouin zone integration, and mesh size of 0.2 Bohr; (iv) 12-atom (3,3) carbon nanotube with the SCAN variant of the meta-GGA exchange-correlation, 10 grid points for Brillouin zone integration, and a mesh size of 0.22 Bohr; and (v) 2-atom primitive cell of germanene with HSE variant of hybrid exchange-correlation, 4$\times$4 grid for Brillouin zone integration, and mesh size of 0.2 Bohr. Unless specified otherwise, we employ ONCV pseudopotentials from the SPMS set. In all cases, we use 12-th order centered finite differences for discretizing the equations, and perform single-point calculations. We present the results so obtained in Fig.~\ref{Fig:Examples}. To verify the accuracy of the results obtained by M-SPARC, we compare them against highly converged results obtained using the established planewave codes ABINIT \cite{gonze2002first} and Quantum Espresso (QE) \cite{giannozzi2009quantum}. It is clear that there is excellent agreement between M-SPARC and ABINIT/QE, verifying the accuracy of the M-SPARC code. Indeed, the agreement further increases on refining the discretization, i.e., choosing smaller values for the mesh-size. 

The new functionalities in M-SPARC v2.0.0 allow for first principles simulations with significantly higher fidelity compared to v1.0.0. Given that some of these features, in particular those involving meta-GGA and hybrid functionals, are noticeably more expensive than standard GGA, new methods/algorithms to accelerate such calculations are highly desired, for which M-SPARC provides a convenient avenue for rapid prototyping.

\section*{Acknowledgements}
\label{acknowledgements}
This work was supported by grant DE-SC0019410 funded by the U.S. Department of Energy, Office of Science. The views and conclusions contained in this document are those of the authors and should not be interpreted as representing the official policies, either expressed or implied, of the Department of Energy, or the U.S. Government.



\bibliographystyle{elsarticle-num} 


\end{document}